\documentclass[dvips,12pt]{article}
\usepackage{epsfig,rotating,pifont,fancybox,cite}
\def\be{\begin{equation}}
\def\ee{\end{equation}}
\def\barr{\begin{array}}
\def\earr{\end{array}}
\def\ra{\rightarrow}
\def\dis{\displaystyle}
\def\simlt{\stackrel{<}{{}_\sim}}

\catcode`@=11 
\def \gsim{\mathrel{\mathpalette\@versim>}}
\def \lsim{\mathrel{\mathpalette\@versim<}}
\def \@versim#1#2{\lower0.4ex\vbox{\baselineskip\z@skip\lineskip\z@skip
     \lineskiplimit\z@\ialign{$\m@th#1\hfil##\hfil$%
     \crcr#2\crcr\sim\crcr}}}
\catcode`@=12 
\title{
\vspace*{-1.3cm}
\begin{flushright}
\normalsize{
ANL-HEP-PR-01-085 \\
MRI-P-010901\\
EFI-2001-37}
\end{flushright}
\vspace{1.cm}
\Large
\textbf{ \Huge Beautiful Mirrors and Precision Electroweak Data
}
\vspace*{1.cm}
\author{\large\textbf{D. Choudhury$^{a,b}$}, \textbf{T.M.P. Tait$^a$} and
\textbf{C.E.M.~Wagner$^{a,c}$}\\ \\
$^a$\normalsize\emph{HEP Division, Argonne National Laboratory,
9700 Cass Ave.,
Argonne, IL 60439, USA} \\
$^b$\normalsize\emph{Harish-Chandra Research Institute, Chhatnag Road,
Jhusi, Allahabad 211 019, India} \\
$^c$\normalsize\emph{Enrico Fermi Institute, Univ. of Chicago, 5640
Ellis Ave., Chicago, IL 60637, USA}}}

\begin{document}
\maketitle
\vspace*{1cm}
\begin{abstract}
The Standard Model (SM) with a light Higgs boson provides a very good 
description
of the  precision electroweak observable data coming from the LEP, SLD and
Tevatron  experiments. Most of the observables, with the notable exception
of the forward-backward asymmetry of the bottom quark,
point towards a Higgs mass far below its current experimental bound.
The disagreement, within the SM, 
between the values for the weak mixing angle as obtained from the
measurement of the leptonic and hadronic asymmetries at lepton colliders,
may be taken to indicate 
new physics contributions to the
precision electroweak observables. In this
article we investigate the possibility 
that the  inclusion of additional bottom-like quarks 
could help resolve this discrepancy. 
Two inequivalent assignments for these new quarks 
are analysed. The resultant fits to the electroweak data
show a significant improvement when compared to that 
obtained in the SM.
While in one of the examples analyzed, the exotic
quarks are predicted to be light, with masses below 300 GeV, and the
Higgs tends to be heavy,
in the second one the Higgs is predicted to be light, with a mass
below 250 GeV,  while the quarks tend to be heavy, with masses of about 
800 GeV.
The collider signatures associated with the new exotic quarks, as well as the
question of unification of couplings within these models and a 
possible cosmological implication of the new physical degrees of freedom
at the weak scale are also discussed.
\end{abstract}
\thispagestyle{empty}
\newpage
\section{Introduction}

The electroweak precision tests, driven primarily by the
experiments at LEP, the Tevatron and the SLC, have, in recent years,
held much of the attention of the field.
Taken in conjunction with the measurement of the top mass
and certain other low energy measurements, these
experiments have vindicated the Standard
Model (SM) to an unprecedented degree of accuracy~\cite{EWWG}.
While startling deviations
from the SM expectations have occasionally 
appeared, 
only to disappear later as the precision increased, the results of the
precision tests have been remarkably steady over the last
five years. Yet, certain discrepancies persist. It is thus contingent upon
us to examine their significance and especially to ascertain whether
they could be pointers to new physics at the weak scale.

In this article, we shall concentrate upon the most obvious
of such a possible deviation~\cite{Chanowitz:2001bv}, 
namely the forward-backward asymmetry
($A^b_{FB}$) of the $b$-quark, the measurement of which shows a $2.9 \sigma$
deviation from the value predicted by the best fit to the precision
electroweak observables within the
SM~\cite{EWWG,LanEr}.
One might, of course, argue that this discrepancy is but a
result of experimental inaccuracies and/or just a large statistical
fluctuation. This viewpoint is supported, to some extent, by
the observation that the corresponding SLD measurement of the
$b$-asymmetry factor $A_b$ 
using the LR polarized $b$ asymmetry is in much better agreement
with the SM~\cite{EWWG}. It has also been argued that any
correction to the $\bar b b Z$ vertex, large enough to `explain'
$A^b_{FB}$ would have shown up in the very accurate
measurement of $R_b$, the branching fraction of the $Z$ into $b$'s.
However, we shall demonstrate that this need not be so. But
more importantly, given 
the remarkable consistency amongst the four LEP experiments
as regards $A^b_{FB}$, it is perhaps worthwhile to take this deviation
from the SM seriously and to speculate on possible explanations
thereof.

Let us begin by reviewing the relevant data at the $Z$-peak.
We parametrize the effective $Zb \bar{b}$ interaction by
\begin{equation}
{\cal L}_{Zb\bar b} = {-e\over s_W c_W} Z_\mu \bar b \gamma^\mu
\left[{\bar g}^b_L P_L + {\bar g}^b_R P_R \right] b
	\label{Lagrangian}
\end{equation}
where $s_W\equiv\sin\theta_W$, $c_W\equiv\cos\theta_W$
and $P_{L, R}$ are the chiral projection operators. An analogous
definition holds for the other fermions. Within the SM, the
tree-level values of the chiral couplings $g^f_{L, R}$ are determined
by gauge invariance. The weak radiative corrections to the same
are well-documented and are insignificant for all but the $b$-quark.
Clearly then,
\begin{equation}
R_b \equiv \frac{\Gamma (Z \ra b \bar b)}{\Gamma (Z \ra {\rm hadrons})}
	\simeq \frac{ ({\bar g}^{b}_L)^2 + ({\bar g}^{b}_R)^2 }
	       { \sum_q \left[ ({\bar g}^{q}_L)^2 
				+ ({\bar g}^{q}_R )^2 \right] }
	\label{rb}
\end{equation}
where the sum is to be done over all the light quarks.
The forward-backward asymmetry at LEP, on the other hand, is
given by
\begin{eqnarray}
 A^b_{FB} |_{\sqrt{s} \simeq m_Z} & = & \frac{3}{4} \; A_\ell \; A_b
    \label{a_fb}
\end{eqnarray}
with
\begin{eqnarray}
A_b & \simeq &	\frac{ ({\bar g}^{b}_L)^2 - ({\bar g}^{b}_R)^2 }
	     { ({\bar g}^{b}_L)^2 + ({\bar g}^{b}_R)^2 }  \;
\nonumber\\
A_\ell  & \simeq &        \frac{ (g^{\ell}_L)^2 - (g^{\ell}_R)^2 }
	     { (g^{\ell}_L)^2 + (g^{\ell}_R)^2 } .
\end{eqnarray}
Small corrections also accrue to  the above observable from
a non-zero $b$-quark and $c$-quark
masses as well as QCD, electroweak and electromagnetic vertex
corrections~\cite{Chetyrkin:1997hm,Degrassi:1991ec,Larin:1995va}.
Whereas the observed values are
\begin{equation}
	R_b (obs) = 0.21646 \pm 0.00065 \ , \quad
	A^b_{FB} (obs)  = 0.0990 \pm 0.0017 \ ,
    \label{expvalues}
\end{equation}
the SM expectations for a top quark mass of 174.3 GeV and a Higgs
mass close to its present experimental bound, 
are $R_b (SM) \simeq 0.2157$ and $A^b_{FB} (SM) \simeq 0.1036$.
Thus, while the observed value for $R_b$ is consistent with the
SM, that for $A^b_{FB}$ shows, as emphasized before,
a relatively large deviation from the predicted value.
This relatively large discrepancy may
be reduced by choosing larger Higgs masses, although only 
at the cost of worsening
the agreement between theory and experiment for other observables,
most notably the lepton asymmetries.

It has been noted, for example in Ref.~\cite{Altarelli:2001wx},
that the overall consistency of the SM with the data improves 
if we dismiss altogether the measurement 
of the forward-backward asymmetry.
Such an act of exclusion leads to a preference for new
physics scenarios that produce a negative shift in the oblique
electroweak parameter $S$~\cite{Peskin:1992sw},
an example being provided by 
supersymmetric theories with light sleptons~\cite{Altarelli:2001wx}.
We, instead,
choose to consider all experimental data on equal footing. 

In this article, we investigate a possible way
of resolving the disagreement between the hadronic and leptonic
asymmetries through the introduction of 
new quark degrees of freedom at
the weak scale thereby inducing non-trivial 
mixings with the third generation of quarks.
In section 2, we examine the experimental status in order to 
determine the necessary modifications 
in the couplings of the right- and left-handed bottom quarks.
As the required modification in the right-handed sector
turns out to be 
too large to be obtainable via radiative corrections, we
investigate, in section 3, the possibility that tree-level mixing
of the bottom quark with exotic quarks might be 
responsible for the observed deviations. All possible 
assignments for such quarks are examined for their 
effects on the precision electroweak observables and 
the two simplest choices identified.
The fits to the data for the two cases
are presented in sections 4 and 5 respectively.
Other phenomenological consequences, including the question of
unification, will be investigated in sections 6 and 7. We reserve
section 8 for our conclusions.

\section{Bottom Quark Couplings Confront Data}

Let us assume a purely phenomenological stance and attempt
to determine $\bar g^b_{L, R}$ from the data. Even in the limit of
infinite precision, the ellipse and the straight lines representing the
solution spaces for eqs.(\ref{rb}, \ref{a_fb}) intersect at {\em four}
points with the coordinates given by
\begin{equation}
(\bar g_L^b, \bar g_R^b) \approx (\pm 0.992 g_L^b(SM), \pm 1.26 g_R^b(SM) ) \ ,
\label{coupnum}
\end{equation}
where we indicate on the right the approximate values of the left- and right-
handed couplings necessary to fit the bottom-quark
production data at the Z-peak\footnote{A similar analysis, although
restricted to modifying the magnitude but not the sign of the couplings,
was performed in Ref.~\cite{Haber:2000zh}}.
Clearly, no experiment performed at
the $Z$-peak can reduce the degeneracy any further.

\begin{figure}[htb]
\vspace*{-5cm}
\centerline{
\epsfxsize=12cm\epsfysize=15cm
                     \epsfbox{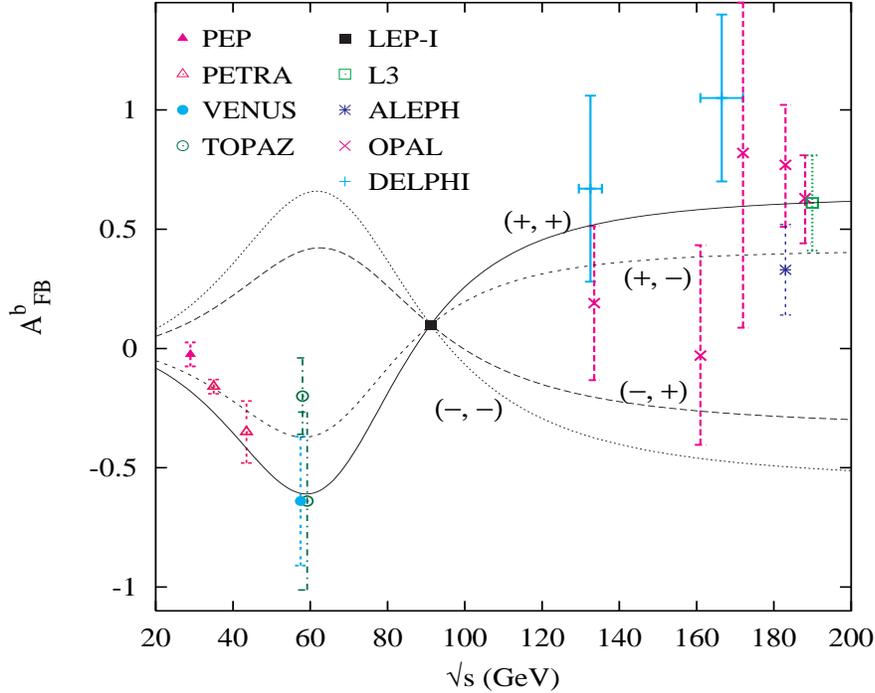}
}
\caption{\em The forward-backward asymmetry for the $b$-quark
	as a function of $\sqrt{s}$ for the four solutions
	of eq.(\protect\ref{coupnum}). 
	The signs in the parentheses refer to those 	
	for ($\bar g_L^b, \bar g_R^b)$ in the 
	same order as in eq.(\protect\ref{coupnum}) with 
	$(+, +)$ being SM-like. 
	The experimental data correspond to the measurements
	reported in 
	Refs.\protect\cite{DELPHI, ALEPH, L3, OPAL, inoue, shimonaka,
	nakano, VENUS, PEP, PETRA_jade, PETRA_tasso}.
	}
\label{fig:afb_rts}
\end{figure}
Off the $Z$-peak though, the photon-mediated diagram becomes important
thereby affecting the forward-backward asymmetry of the bottom-quark.
Such data, thus, could discriminate amongst the four 
solutions described above. 
The asymmetry is easy to calculate
and in Fig.~\ref{fig:afb_rts}, we plot the same as a
function of the center of mass energy of the $e^+ e^-$ system
for each of the 
solutions\footnote{Had we instead held the magnitudes of the couplings 
	to their SM values, the resulting curves would have been 
	barely distinguishable from those in 
	Fig.~\protect\ref{fig:afb_rts}.}
 in eq.(\ref{coupnum}).
It is quite apparent that the two solutions with
$\bar g_L^b \approx - g_L^b (SM)$ can be summarily discarded.
Interestingly enough, the data does not readily discriminate between the
two remaining solutions. This, though,  is not unexpected as
$ |g_R^b | \ll |g_L^b|$ within the SM. A similar analysis can be performed
for $R_b$ as well, but the off-peak measurements of this variable are
not accurate enough to permit a similar level of discrimination.

It is quite interesting to note that the agreement with
the next best measurement of $A^b_{FB}$, viz. that at {\sc petra} (35 GeV)
is much better for the $(+, -)$ choice than for the 
SM (or the `SM-like' solution). This 
observation can be quantified by performing a $\chi^2$ test including 
all the data shown in Fig.\ref{fig:afb_rts}. It can easily be ascertained 
that the $\chi^2$ is indeed significantly improved if the sign of $\bar g^b_R$ 
were to be reversed. Whether this information actually calls for a 
such a reversal is, of course, open to interpretation. 

\begin{figure}[htb]
\vspace*{-3cm}
\centerline{
\epsfxsize=8cm\epsfysize=10.0cm
                     \epsfbox{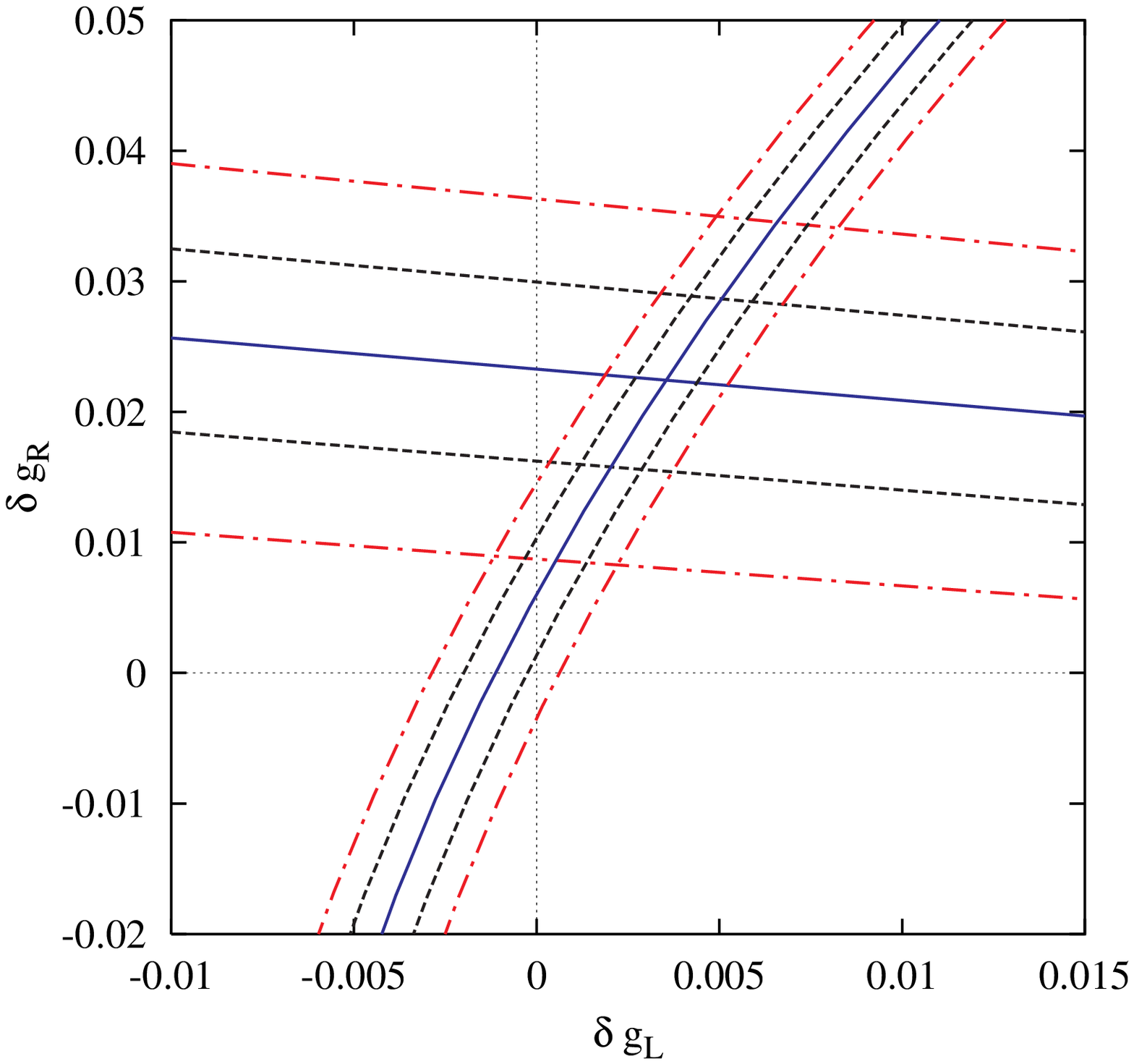}
        \hspace*{-3ex}
\epsfxsize=8cm\epsfysize=10.0cm
                     \epsfbox{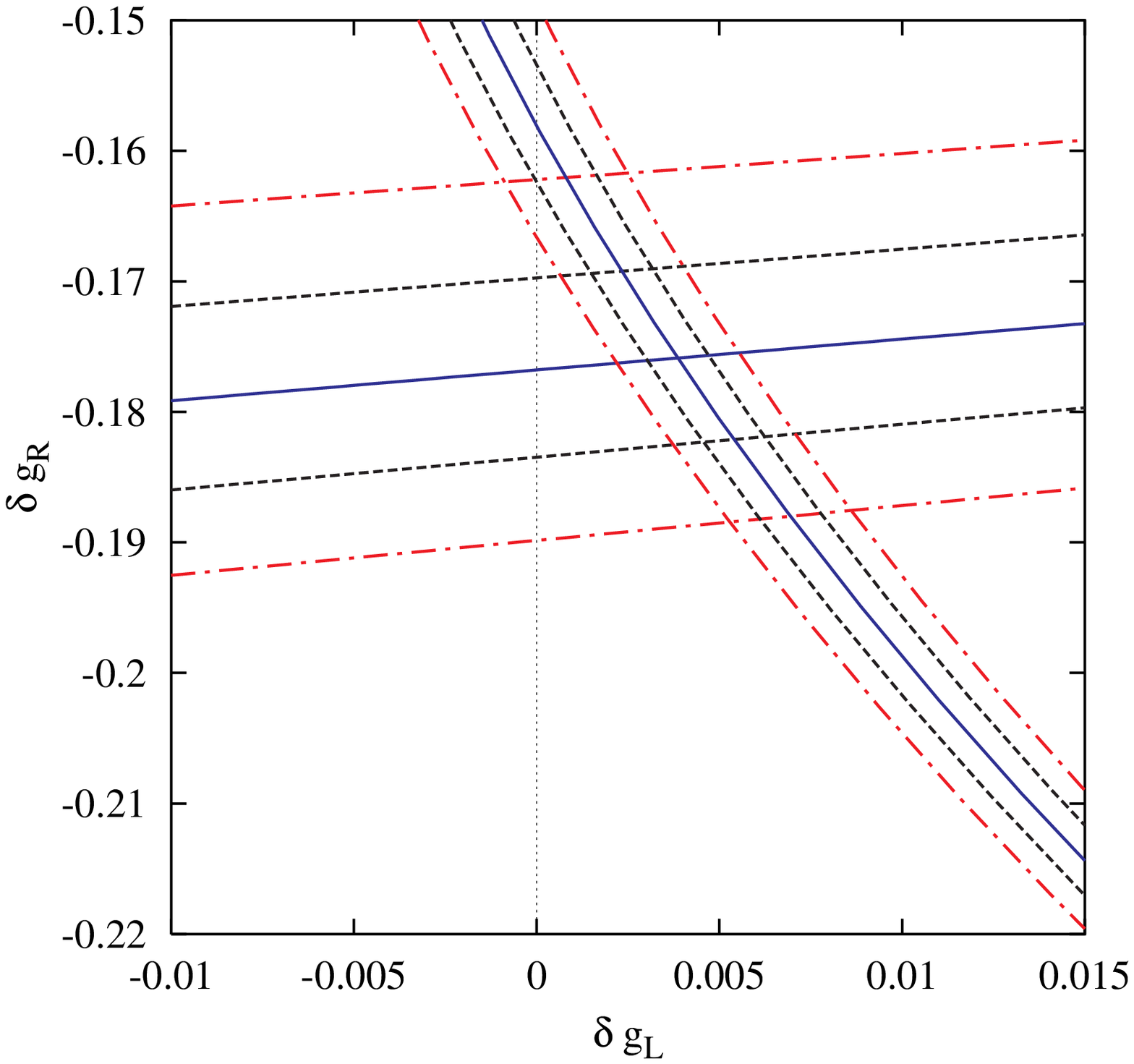}
}
\caption{\em The regions in the $Z\bar b b $ coupling parameter space
	that are favoured by the observed values of $A^b_{FB}$ (flatter
	curves) and $R_b$ (steeper curves).
	For each set, the innermost curve leads to the experimental
	central value while the sidebands correspond to the
	$1 \sigma$ and $2 \sigma$ error bars. The Standard Model
	point is at the origin.
	}
\label{fig:param}
\end{figure}

Any resolution of the $A^b_{FB}$ anomaly through a modification of the
$Z b \bar b$ couplings must then lie within one of two disjoint
regions of the parameter space, regions that we exhibit in
Fig.~\ref{fig:param}. What immediately catches the eye is that
the required shifts in the coupling satisfy
$| \delta g_R| \gg | \delta g_L|$, a condition that would prove crucial
at a later stage of our analysis.  At this point, it is perhaps worthwhile
to note that the two other (ruled out) branches of the solution space
would have required a very large $| \delta g_L|$, a shift that is
very hard to obtain in any reasonable model.

\section{Beautiful Mirrors}

We now turn to the question of whether the
required $\delta g_{R, L}$ could arise naturally
as consequences of ordinary-exotic quark mixing.
To keep the discussion simple, yet without losing track
of any subtle effects, let us, for now, confine ourselves to just
one additional set of quarks. Any extension of the model would 
not change the qualitative aspects of our analysis.
We shall also, for the time being, neglect any mixing with quarks
of the first two generations\footnote{A negligibly small mixing
with the first and second generations may be enforced, for instance,
by additional gauge interactions, such as 
Top-color~\cite{Hill:1991at, Bonisch:1991vd, Hill:1995hp, Dobrescu:1998nm,
Chivukula:1999wd}, 
Top-flavor\cite{Chivukula:1995qw, Malkawi:1996fs, He:2000vp, Muller:1996dj} 
or Bottom-color~\cite{Popovic:2001bn}.}.
At this stage, we do not make any
further assumptions about the quantum numbers of these
new quarks. Working in the basis $(b_1', b_2')$, where the primes
indicate weak-interaction eigenstates and $b_2'$ refers to the
exotic $b$-quark, the mass matrix can be parametrized as
\begin{equation}
{\cal L}_{m_b} = - \sum_{i j} \bar b_{iL}' M_{ij} b_{jR}' \; + {\rm h.c.} ,
\quad M \equiv
 \pmatrix{ M_{11} & M_{12} \cr
                        M_{21} & M_{22}
\cr }
\end{equation}
where the subscripts $L, R$ refer to the quark chirality, and
the (in general, complex) elements $M_{ij}$ represent either
a bare mass term or one derived from the Higgs mechanism.
It is a straightforward task, then, to obtain the mixing matrices
for the left- and right-handed quarks (as well as the mass
eigenvalues) by diagonalizing the matrices $M M^\dagger$ and
$M^\dagger M$ respectively. 
%
Ordinary-exotic mixings would generically introduce
additional parameters in the charged current structure
and, more importantly for us, in the neutral current
sector as well. Any such deviation from the SM structure
depends crucially on the isospins of the exotics, and
for the new left- and right-handed $b'$ fields, we denote
these values by $t_{3 L(R)}$.
Let us concentrate on the neutral currents in the $b$-sector,
more specifically on the part
independent of the charge generator\footnote{Any
mixing which respects $U(1)_{\rm em}$ must be between objects of the
same charge $Q$ and thus there are no mixing effects proportional to $Q$.}
Expressed in  terms of the physical states (mass eigenstates),
these can be parametrized as
\begin{equation}
\begin{array}{rcl}
J_\mu^3 (b) & = &  \dis \frac{e}{s_W c_W}\dis
	\sum_{i j} \bar b_i \gamma_\mu
			(L_{i j} P_L + R_{i j} P_R) b_j \ ,
   \\
\nonumber\\
L & \equiv & \dis
	\pmatrix{t_{3L} s_L^2 - \frac{1}{2}c_L^2 &
			- (t_{3 L} + \frac{1}{2}) s_L c_L \cr
                    - (t_{3 L} + \frac{1}{2}) s_L c_L   &
			t_{3L} c_L^2 - \frac{1}{2}s_L^2 \cr}
	\\[2ex]
\nonumber\\
R & \equiv & \dis
	 \pmatrix{t_{3R} s_R^2 &
			- t_{3 R} s_R c_R \cr
                    - t_{3 R} s_R c_R   &
			t_{3R} c_R^2 \cr}
\end{array}
	\label{neut_current}
\end{equation}
where $s_{L, R} \equiv \sin \theta_{L, R}$ etc parametrize the left- and
right-handed mixing matrices in the $b$-sector and $s_W$ ($c_W$)
denote the sine (cosine) of the weak mixing angle. As expected, we would
have flavour-changing neutral currents if either $t_{3 L} \neq - 1/ 2$
or $t_{3 R} \neq 0$. The presence of any such coupling would play a
crucial role in the discovery of such an exotic and
we shall return to this later. Since the shifts in $g^b_{L, R}$ are given by
\begin{equation}
	\delta g^b_L = \left( t_{3 L} + \frac{1}{2} \right) s_L^2 \ ,
	\qquad
	\delta g^b_R = t_{3 R} s_R^2  \ , 
	\label{delta_g}
\end{equation}
it follows that the right handed component
of the exotic cannot be a $SU(2)_L$ singlet.

In principle, we could allow each of $b'_L$ and $b'_R$ to lie
in any (and inequivalent) representation of
$SU(3)_c \otimes SU(2)_L \otimes U(1)_Y$.
However, the requirement of anomaly cancellation indicates that
a vector-like coupling for the exotics is the most economic choice.
In addition, the introduction of
vector-like fermions, unlike the one of their chiral counterparts,
do not lead  to large contributions to the oblique electroweak parameter
$S$~\cite{Peskin:1992sw}, thereby preserving the agreement with precision
data. We shall, therefore, limit ourselves to only such quarks. We 
refer to these exotic quarks as (beautiful) mirrors in the sense that 
they occur in vector-like pairs and they have the same electric and color
charges as the ordinary bottom- (or beauty-) quark.

As we have already mentioned, nonzero quark mixing requires that there
be mass terms connecting the ordinary $b$-quark to its exotic counterpart.
Demanding that the only scalars in the theory be the
$SU(2)$ Higgs boson doublets
restricts the choice of the exotics to
a $SU(2)$ singlet and two varieties each of $SU(2)$ doublets and triplets.
The phenomenological requirement of $t_{3R} \neq 0$ eliminates the
singlet and one of the triplets as the possible source for the large
modification of the right-handed bottom quark coupling.
The choice then devolves to one of
$\Psi_{L, R} = (3, 2, 1/6), (3, 2, -5/6)$ and $(3, 3, 2/3)$. The
phenomenological consequences of the first and last possibilities
are similar and hence we shall concentrate on studying the first
two cases.

\section{Scenario with Standard Mirror Quark Doublets}

Our first scenario relies on the introduction of the fermion doublets
\begin{equation}
\Psi_{L,R}^T = (\chi,\omega) \equiv (3,2,1/6) ,
\end{equation}
which are
a mirror copy of the standard quark doublets of the Standard Model.
The most general Yukawa and mass term in the Lagrangian is then
\begin{equation}
  {\cal L} \supset  - \left( y_1 \overline{Q'_L} + y_2 \overline{\Psi_L}
			\right) b'_R \phi
                  - \left( x_1 \overline{Q'_L}
+ x_2 \overline{\Psi_L^{'}} \right)
				t'_R \tilde \phi
		  - M_1 \overline{\Psi_L^{'}} \Psi_R^{'}
+ h.c., 
\end{equation}
where the primes, once again, denote weak eigenstates. Note that,
on account of $\Psi_L^{'}$ and $Q'_L$ having the same quantum numbers, a
mass term of the form $\overline{Q'_L} \Psi_R$ can be trivially rotated
away. In the basis $(b',\omega')$,
we then have a mass matrix of the form
\begin{equation}
	M_b = \pmatrix{Y_1 & 0 \cr
		       Y_2 & M_1 \cr} \ , \quad Y_i \equiv y_i
						\langle \phi \rangle
	\label{matrix_in_model}
\end{equation}
and an analogous one for the top. For the sake of 
simplicity, we shall assume that
the mass matrices are real. In the phenomenologically interesting
regime of $Y_1 \ll Y_2 < M_1$, we have, for the eigenvalues of the
mass eigenstates $b,\omega$ and the
mixing angles
\be
\barr{rclcrcl}
    m_b& \approx & \dis Y_1 \left(1 + \frac{Y_2^2}{M_1^2} \right)^{-1/2} \ , 
	& \quad & 
    m_{\omega} & \approx & \left(M_1^2 + Y_2^2  \right)^{1/2} \ ,
\\[2.2ex]
    \tan \theta_R^b & \approx & \dis \frac{-Y_2}{M_1} \ & \quad &
    \tan \theta_L^b & \approx & \dis \frac{-Y_1 Y_2}{M_1^2 + Y_2^2} \ ,
\earr
\ee
with analogous expressions for the top-sector. A few points
are to be noted:
\begin{itemize}
\item[$\bullet$]Since both  $\omega'_L$ and $\chi'_L$ have the same quantum
	numbers as their ordinary counterparts, neutral currents
	in these sectors remain unmodified.
\item[$\bullet$] As $\delta g_R^b < 0$ (while $g_R^b(SM) > 0$), a small
	$\delta g_R^b$ would only worsen the fit. Rather, we must
	demand a large negative correction that would take us to the
	second allowed region in the parameter space 
	(see Fig.\ref{fig:param}). For example,
	a $1 \sigma$  agreement for each of $A^b_{FB}$ 
        and $R_b$ is obtained for
	\be
	\delta g_R^b = \frac{- s_R^2}{2} \approx -0.165
		\quad \Longrightarrow \quad
	 Y_2 \approx 0.7 \: M_1 .
		\label{solution}
	\ee

\item[$\bullet$] The top-sector mass matrix is as in eq.(\ref{matrix_in_model})
	but with $y_i \ra x_i$. Since the $y$'s and $x$'s are independent, one
	could, in principle, set $x_2 = 0$ (this, for example, could
	be ensured by imposing a discrete symmetry).  In such a case,
	the top sector sees no additional mixing and $x_1$ is the usual
	top Yukawa coupling.

\item[$\bullet$] Since no  exotic quark has yet been seen 
	at the Tevatron collider,   $M_{1} \gsim 200~{\rm GeV}$.

\item[$\bullet$] In  general, due to the large
        mixing in the bottom sector and the fact that the right-handed
        mirror quarks carry non-trivial weak charges, 
        a potentially large correction to the
	precision electroweak parameters will accrue.
\item[$\bullet$] A right-handed $W$-$t$-$b$ interaction is induced
with strength proportional to $s_R^b s_R^t$. Measurement of $b \to s \gamma$
requires $s_R^b s_R^t < 0.02$~\cite{Larios:1999au}, 
leading us to consider the case of
negligible $\chi$-$t$ mixing.
\end{itemize}

In order to address the question of how well does this scenario fit
the data, we have computed the corrections to the $S$, $T$ and $U$
parameters, with respect to a reference Higgs mass value of 115 GeV.
While the corrections to $U$ are small, the  corrections
to $T$ and $S$ are large and increase with the overall scale of quark masses.
For instance, for $M =$ 200, 225, 250 GeV the corrections to the
$T$ parameter are $\Delta T \simeq$ 0.35, 0.42, 0.54 respectively, while
the correction to the parameter $S$ is somewhat insensitive to the 
masses and measures $\Delta S \simeq 0.1$.

The large corrections to the $T$ parameter, together with the relatively
large corrections to the right-handed bottom couplings, tend to increase
the hadronic width and the total width of the Z to unacceptable levels.
This problem can be ameliorated by including the mixing of the
bottom quark with a quark $\xi_{R,L}$
carrying the quantum numbers of the right-handed
bottom quark and its mirror partner. In the basis $(b',\omega',\xi')$,
the simplest modification to the mass matrix that fulfills
this requirement is given by:
\begin{equation}
	M_b = \pmatrix{Y_1 & 0 & Y_3\cr
		       Y_2 & M_1 & 0\cr
                       0   & 0   & M_2 } \ , \quad Y_i \equiv y_i
						\langle \phi \rangle
	\label{M_b_singlet}
\end{equation}
Note that, as happens with $(M_b)_{12}$, the element $(M_b)_{31}$
could also be trivially rotated away. The inclusion of small,
but non-zero, values of the elements $(M_b)_{23}$ and $(M_b)_{32}$
only serves to complicate matters without modifying the main phenomenological
consequences of this model. 

Ignoring small terms proportional to the bottom quark mass, the
left-handed mixing angle is now given by
\begin{equation}
s_L \simeq \frac{Y_3}{\sqrt{Y_3^2 + M_2^2}}
\end{equation}
The main effect of the mixing with these weak singlet quarks is to
reduce the left-handed coupling of the bottom quark and thus the
partial width of the $Z$ into $b$'s 
and hence into hadrons as such. 
The scenario described above can thus clearly improve the agreement with
$A^b_{FB}$.  
For small values of $s_L$, as demanded by
experimental results, the oblique corrections to the precision electroweak
observables are still dominated by the large mixing of the
bottom quark with the weak mirror quark doublet. 
The presence of the new quarks will, of course, induce additional 
radiative corrections to the $b$-quark couplings. It is easy to see 
though that,  given the above-mentioned mass and mixing angle 
pattern, these corrections are tiny compared to those induced 
at the tree level, and hence could be safely neglected.
Once again, non-observation of an 
exotic quark at the Tevatron implies $M_2 \gsim 200~{\rm GeV}$.

The parameters $S$, $T$ and $U$ are in one to one correspondence with
the variations of the parameters $\epsilon_3$, $\epsilon_1$ and
$\epsilon_2$, introduced in  Ref.~\cite{Altarelli:1991zd}
with respect to
a given value of the top quark mass and the Higgs mass. The relation
between these parameters is given by:
\begin{eqnarray}
\Delta \epsilon_1  & = &  \alpha T , 
\quad   \;\;\;\;\;\;\;
\Delta \epsilon_3 = \frac{\alpha S}{4 s^2_W} ,
\nonumber\\
\Delta \epsilon_2 & = & -\frac{\alpha U}{4 s^2_W} , 
\end{eqnarray}
and, within the SM, for a top quark mass $m_t = 174.3$ GeV and a Higgs
mass $m_H = 115$ GeV, $\epsilon_1 = 5.6 \; 10^{-3}$,
$\epsilon_2 = -7.4 \; 10^{-3}$
and $\epsilon_3 = 5.4 \; 10^{-3}$~\cite{Altarelli:2001wx}. 
The dependence of the $\epsilon_i$
parameters on the Higgs and top quark masses may be found, for instance,
in Ref.~\cite{Altarelli:1998et}.
The dependence of the most important observables on 
the parameters $\epsilon_i$ are given by
\begin{eqnarray}
\Gamma_Z & \simeq & 2.489
\; (1 + 1.35 \; \epsilon_1 - 0.46 \; \epsilon_3 + ...) \; {\rm GeV}
\nonumber\\
\sin^2\theta^{\rm eff}_l & \simeq & 0.2310
\; (1 - 1.88 \; \epsilon_3 + 1.45 \; \epsilon_1)
\nonumber\\
\frac{m_W^2}{m_Z^2} & \simeq & 0.7689 \; (1 + 1.43 \; \epsilon_1
- \epsilon_2 -
0.86 \; \epsilon_3) \ ,
\end{eqnarray}
where the dots reflect the contributions associated with the variations
of the b-coupling due to radiative corrections and the mixing with the
$b$ quarks. There is, in addition, a dependence 
on the precise value of $\alpha(M_Z)$ and $\alpha_s(M_Z)$. 
In our computations
we have used the central values for the hadronic contribution to
$\alpha(M_Z)$ namely 
$\Delta \alpha_{\rm had}^{(5)} = 0.02761$~\cite{Burkhardt:2001xp}, 
while the strong gauge coupling was allowed 
to float around the central value of 0.118~\cite{Groom:2000in}.

Variations in the parameter $T$ larger than 0.3 tend to induce a
large positive correction to the total Z width, and are therefore
disfavored by the data. Consequently, this  model leads to a better
fit to the data whenever the new quarks are relatively light and
the Higgs is heavy. In general, a heavy Higgs leads to a negative contribution
to $T$ and a positive contribution to $S$, leading to a better
agreement with the total width of the Z. However, the same effects
worsen the agreement of the data with the leptonic asymmetries, which
mainly depend on $\sin^2\theta^{\rm eff}_l$.
Therefore, the model leads to a correlation of the quark
and Higgs masses: The heavier the new quarks, the heavier the
Higgs needs to be. The value of $\alpha_s$ also plays an important role in
this process, since it may lower the total hadronic width without
modifying the leptonic asymmetries.

We have made a fit to the data within this model.
Including the values of $Y_2$, $M_1$, $\alpha_s$, $m_t$, $m_H$
and $s_L$ as variables in the fit (the fit is quite
insensitive to the scale $M_2$, provided it remains below a
a few TeV), we obtain that the best fit to
the data is obtained for the mirror quark mass parameter $M_1$ close
to the present experimental bound on this quantity, while
the preferred values of the Higgs mass are about 300 GeV.
Raising the quark bound to 250 GeV leads to an optimal
value of the Higgs mass of about 850 GeV. The best fit
gives $\alpha_s \simeq 0.116$ and $m_t \simeq 173$ GeV.
For the exotic sector, the corresponding values are
\begin{equation}
Y_2 \simeq 0.71 \; M_1 , \quad \quad M_1 \simeq 200 \: {\rm GeV}
\end{equation}
while
\begin{equation}
s_L^2 \simeq  0.008.
\end{equation}
The best fit to the ratio $Y_2/M_1$ and to $s_L^2$ are virtually
independent of $M_1$ for 200 GeV $\simlt M_1 \simlt$ 250 GeV.

\begin{figure}[htb]
\vspace*{-1.3cm}
\centerline{
\epsfxsize=10cm\epsfysize=11cm
                     \epsfbox{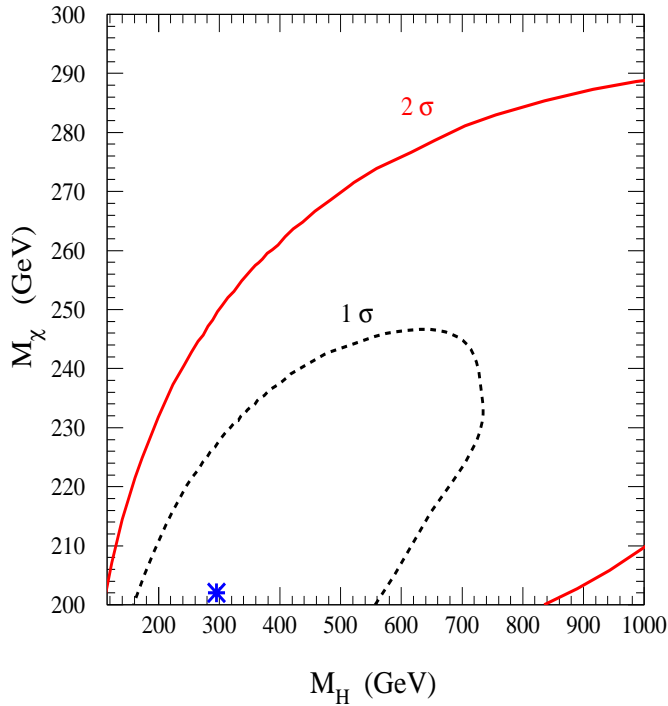}
}
\caption{\em Region in the $m_H$--$m_{\chi}$ parameter space 
	(in the model with standard mirror quark doublets)
	that is consistent with the best fit point (marked)
	at the 68\% C.L. and 99.5\% C.L. respectively.
	}
\label{fig:ellipse1}
\end{figure}

In Fig.~\ref{fig:ellipse1} we show the 1- and 2-$\sigma$ 
regions in the $m_H$--$m_{\chi}$ parameter space determined by the
best fit to the data. As emphasized above, the model leads to
a preference for light quarks, with masses below or about 250 GeV,
within the reach of the Tevatron collider (see section 6), 
while the preferred values of the Higgs mass are much larger than
in the Standard Model, a feature that appears in many 
models~\cite{Peskin:2001rw}. 

For the parameters providing the best fit, all measured precision 
electroweak observables,
including the lepton and hadron asymmetries and the Z widths
are within $2 \sigma$ of the predictions of this model, and, in
particular, the bottom asymmetries are within $1 \sigma$ of the
predicted values. Similar
results are obtained for slightly larger values of $M_1$, although
the model is clearly disfavored for quark masses $M_1 > 250$ GeV.
While the agreement between theory and experiment for the 
bottom-quark asymmetries is remarkably better than
in the Standard Model, the lepton asymmetries remain essentially 
the same as in the SM.
This is due to the tension between these observables and the
total Z-width within this model and is reflected in the fact that
the fit produces
\begin{equation}
\sin^2\theta^{\rm eff}_l \simeq 0.2315 \ .
\end{equation}
As a result, the left-right lepton asymmetry is about
1.9 standard deviations away from the value measured at SLD
thus marginally worsening the discrepancy obtained within
the Standard Model.  The $W$ mass is, instead,  in excellent agreement with
the predictions of this model.

\section{Top-less Mirror Quark Doublets}

Let us now analyze the case in which the mirror quarks belong to
a doublet in which there is no quark with the same charge as the
top quark, viz., 
\begin{equation}
\Psi_{L,R}^T = (\omega,\chi) \equiv
(3,2,-5/6).
\end{equation}
This  model has some advantages with respect to the model
analyzed above. First of all, since the weak partner of the
$\omega$ has charge $-4/3$, there is no mixing involving the
top quark. Second, the model allows for a modification of the
right-handed bottom couplings with moderate mixing angles.
In the basis ($b',\omega'$), the mass matrix reads
\begin{equation}
	M_b = \pmatrix{Y_1 & 0 \cr
		       Y_R & M_1 \cr} \ , \quad Y_i \equiv y_i
						\langle \phi \rangle \ ,
\end{equation}
where the zero entry is now enforced by gauge invariance. The
right and left-handed mixing angles have similar expressions
to the ones found in the above model. However, the mixing of the
right-handed bottom leads to a positive shift of $g_R^b$,
\begin{equation}
\delta g_R^b \simeq \frac{1}{2} s_R^2
\end{equation}
and therefore small values of $s_R$ can lead to a relevant shift
of $A_{FB}^b$ in the direction required by experiment.

As in the previously analyzed model, the values of $R_b$ and of the
hadronic width can be improved by allowing an additional mixing
with a quark $\xi$ with the same quantum numbers as the right-handed
bottom quark and its mirror partner. In the basis ($b',\omega',\xi'$),
we assume a mass matrix quite similar to the one in the last section:
\begin{equation}
	M_b = \pmatrix{Y_1 & 0 & Y_L\cr
		       Y_R & M_1 & 0\cr
                       0   & 0   & M_2 } \ , \quad Y_i \equiv y_i
						\langle \phi \rangle
\end{equation}
As with the previous model, the matrix element $(M_b)_{31}$ can
be trivially rotated away, while the inclusion of small (compared to 
$M_i$) but non-vanishing
matrix elements $(M_b)_{23}$ and $(M_b)_{32}$ do not change the 
main results of our analysis.

Ignoring small effects induced by the bottom mass, the left-handed
and right-handed mixing angles are given by
\begin{equation}
s_L \simeq \frac{Y_L}{\sqrt{Y_L^2 + M_2^2}}, \;\;\;\; \quad
s_R \simeq \frac{Y_R}{\sqrt{Y_R^2 + M_1^2}}.
\end{equation}
The main difference between this model and the one analyzed before
lies in the smallness of all the mixing angles. Since all
Yukawa couplings are small compared to the explicitly gauge invariant
masses, the corrections to the oblique parameters $S$, $T$ and $U$
are small. The corrections to $T$ become relevant only for quark
masses above 500 GeV, while the corrections to $S$ and $U$ remain
small even for masses in the multi-TeV range. Since the 
expected bottom-quark
asymmetry has now migrated much closer 
to the measured value, the data now prefers non-negligible values
of the $T$ parameter so as to permit a better agreement with the
lepton asymmetries and the $W$ mass. This can only be achieved
by pushing the quark masses up, while keeping the Higgs mass
close to its experimental lower bound.

We have analyzed all the precision observables in the 
context of this model, as described by the parameters
$m_H$, $m_t$, $\alpha_s$, $M_1$,
$Y_R$, $M_2$ and $Y_L$.
The best fit to the data is obtained for a Higgs
mass close to the present experimental bound and mirror quark
doublets with mass of about 
\begin{equation}
M_1 \simeq 825 \: {\rm GeV} \quad {\rm and} 
	\quad Y_R \simeq 160 \: {\rm GeV} ,
\end{equation}
implying that $s_R^2 \simeq 0.036$. The best fit value
of $M_2$ is close to its experimental bound, 
$M_2 \simeq 200$ GeV, while $Y_L \simeq 15$ GeV, leading to 
$s_L^2 \simeq 0.006$. 
Similar to the previously analyzed scenario,
changing $M_2$ while keeping the ratio of $Y_L/M_2$ does not
alter the fit in any significant way. The best fit values
of $\alpha_s$ and $m_t$ are $\alpha_s \simeq 0.116$ and
$m_t \simeq 176$ GeV.

\begin{figure}[htb]
\vspace*{-1.3cm}
\centerline{
\epsfxsize=10cm\epsfysize=11cm
                     \epsfbox{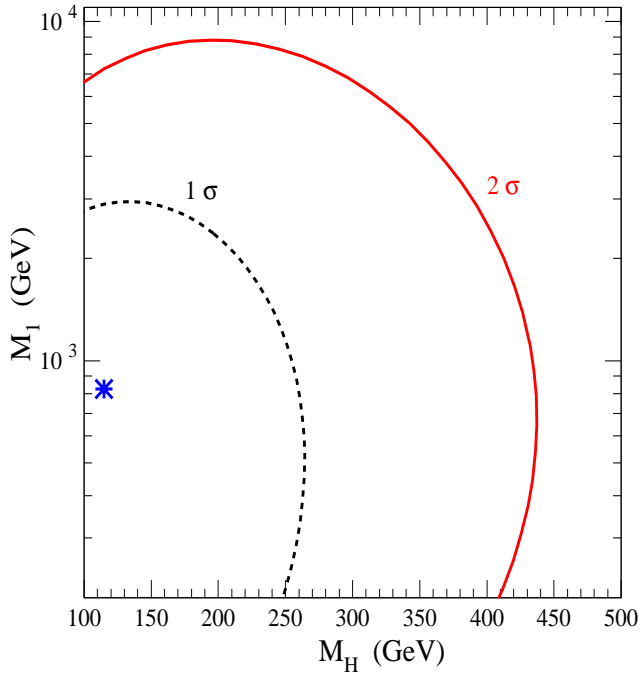}
}
\caption{\em 
        Region in the $m_H$--$m_{\chi}$ parameter space 
	(in the model with top-less mirror quark doublets)
	that is consistent with the best fit point (marked)
	at the 68\% C.L. and 99.5\% C.L. respectively.
	}
\label{fig:ellipse2}
\end{figure}
In Fig.~\ref{fig:ellipse2} we show the 
1- and 2-$\sigma$ regions in the $m_H$--$M_1$
parameter space obtained by the best fit to the data. 
As emphasized above, the Higgs tends to be light, in the
region most accessible to the Tevatron collider and a
$\sqrt{s} = 500$ GeV linear collider. The quarks, instead,
tend to be heavy with masses of about 1 TeV. 
For $M_1$ above a few TeV though, the Yukawa
coupling $Y_R$ needed to improve the fit to the data
becomes large, spoiling the perturbative consistency
of the theory at low energy scales. 

This model provides a surprisingly good agreement with the
experimental data. For the parameters providing the
best fit to the data, the left-right lepton asymmetry
measured at SLD is 1.2 standard deviations from the theoretically
predicted value, while almost all other measured observables are within
1 $\sigma$  of the predictions of this model. The only exceptions 
are the charm forward-backward asymmetry and the total hadronic
cross section measured at LEP, which stay within 2 $\sigma$ of the
measured values. 
As already pointed out, the fitted value 
\begin{equation}
\sin^2\theta^{\rm eff}_l \simeq 0.2313
\end{equation}
exhibits a much better agreement with the leptonic asymmetries
than in the model with Standard Mirror Quarks.

\section{Implications at Present and Future Colliders}

Although the two models presented above share many features, 
there are subtle differences as far as the collider signatures 
are concerned. We shall examine, in some detail,
the scenario containing a top-like quark and then 
point out the differences with the second model. 

Any new quark~\cite{Frampton:2000xi}, with a mass below or about 250 GeV,
as preferred in the model with standard quark doublets, should
be observable at the next run of the Tevatron collider. The
main decay of the $\chi$-quark will be similar to that of the top
quark, namely,
\begin{equation}
\chi \rightarrow b + W^+ .
\end{equation}
Although the $\omega$-quark tends to be somewhat heavier than 
the $\chi$-quark, yet it should be possible to pair-produce it
at the Tevatron collider, especially if $M_1$ turns out to be 
in the (lower) range preferred by the electroweak fit. 
On account of the phase space restrictions, it 
would decay mainly through  the flavor violating channel,
{\em viz.}, 
\begin{equation}
\omega \rightarrow b + Z
\end{equation}
a mode that has already been looked for at the 
Run I of the Tevatron with a resultant lower limit of about
200 GeV on $m_\omega$~\cite{Affolder:2000bs}.
This very same flavor-changing neutral current interaction, 
would, at a next generation linear collider, lead to 
a rather significant cross section for the process
\be
	e^+ + e^- \ra \bar b + \omega, 
\ee
as well as the conjugate state, thereby leading to rather 
striking signatures. 

There is no prediction for the singlet quark mass within this models.
If they were light, in the range accessible to the Tevatron collider,
they will mainly decay via its flavor violating couplings
\begin{equation}
\xi \rightarrow b + Z \ ,
\end{equation}
or, in the event of a non-zero $(M_b)_{23}$, also into
an $\omega$-quark:
\begin{equation}
	\xi \ra \omega + Z .
\end{equation}
Due to its larger center of mass energy, the LHC, 
should be able to test this model for even larger 
values of the $\omega$- and $\chi$- quark masses. 

Finally, we turn to the Higgs. Note that the Yukawa 
coupling matrix is proportional to that in eq.(\ref{M_b_singlet})
with the gauge invariant masses $M_{1, 2}$ switched off. 
This immediately implies that the Yukawa interactions are not 
diagonal in the mass-basis. This is quite crucial especially 
since the coupling $y_2$ is quite large (actually, of the same
order as the top Yukawa in the SM). While the full expressions
for the resultant Yukawas are cumbersome, they simplify 
considerably in the limit of a vanishing $y_3$ to 
\begin{equation}
   {\cal L}_{b w \phi^0} \sim y_2  \;
		(c_R \; \bar{\omega}_L  b_R - 
              s_R \; \bar \omega_L  \omega_R) \; \phi^0
		+ h.c. + {\cal O}(y_1)
\end{equation}
Thus, for $m_{\phi^0} > m_\omega + m_b$, a condition satisfied 
over almost the entirety of the preferred parameter 
space (see Fig.\ref{fig:ellipse1}), 
the Higgs is afforded 
an additional decay mode. And, if the Higgs is more than twice as 
heavy as the $\omega$ (again, true for a very large part of 
the $1\sigma$ preferred area), a further decay channel would open up, although 
with a branching fraction smaller than that for the flavour-changing mode.
While this results in a severe depletion of the `gold-plated'
modes ($\phi^0 \ra Z Z \ra 4 l$), presumably these non-canonical 
channels would lend themselves easily to discovery, especially with 
the $b$- (and lepton-) richness of the final state.

In the top-less mirror quark model, instead,
the quark doublets are predicted to be
heavier. Only the LHC (or a second generation linear collider)
would have enough center of mass energy to
produce the $\omega$ and $\chi $ in this case. The quark 
$\chi$ (with a $-4/3$ charge) will decay into
\begin{equation}
\chi \rightarrow b + W^{-} \ ,
\end{equation}
leading to a top-like signature with a wrong sign $W$.
Similarly, due to phase space restrictions,
\begin{equation}
\omega \rightarrow Z + b
\end{equation}
Since the flavour changing coupling is smaller in this case as compared 
to the previous model, non-diagonal production at a linear 
collider would be somewhat suppressed. Also, with the Higgs 
preferring to be light, and with its couplings 
to the new states not being as large as in the other model, 
Higgs phenomenology remains largely unchanged from the SM. 

As happens in the
model with standard mirror quarks, there is no prediction for
the singlet mirror quark masses. If they are available at the
colliders, they will decay via its flavor violating coupling:
\be
	\xi \rightarrow b + Z \ .
\ee
And since $\xi$ prefers to be lighter than $\omega$, a non-zero 
$(M_b)_{23}$ would, once again open an additional decay channel, viz.,
\be
    \omega \rightarrow \xi + Z \ .
\ee

\section{Unification}

The question of unification within 
these models is an interesting one. It might be argued that this 
is not a pertinent one, 
since we have not detailed any mechanism to keep the 
Higgs boson and vector quarks naturally light. 
The hierarchy of masses may arise through some 
hitherto undiscovered mechanism, which must be related to the 
one leading to the breakdown of the electroweak symmetry, and will probably
require extra gauge  
symmetries~\cite{Hill:1991at, Bonisch:1991vd, Hill:1995hp, Dobrescu:1998nm,
Chivukula:1999wd, Chivukula:1995qw, Malkawi:1996fs, He:2000vp, Muller:1996dj, 
Popovic:2001bn, Babu:1991gp, Babu:1995pd, Babu:1996zv}. The presence of 
any such extra symmetry would certainly alter unification 
in a highly model-dependent way. It is still possible though to 
discuss the issue of unification without delving into the details of 
the implementation of such a symmetry, as long as one assumes that 
the low-energy  spectrum is determined, besides additional
complete SU(5) multiplets~\cite{Babu:1996zv}. This is the approach that 
we adopt in this section. 

We shall proceed with a one-loop analysis, taking 
into account that the possible two-loop effects are of the 
order of the small threshold effects at the Grand Unification scale, 
and hence become strictly relevant only for the construction of
a complete grand unified model, an objective which is beyond the
scope of this article. 

In the model with standard mirror quarks
the beta-function coefficients of the three gauge couplings are
given by
\begin{eqnarray}
b_3 & = & - 11 + \frac{4}{3} n_g + 2
\nonumber\\
b_2 & = & -\frac{22}{3} + \frac{4}{3} n_g + \frac{n_H}{6} + 2
\nonumber\\
b_1 & = & \frac{4}{3} n_g +  \frac{n_H}{10} +\frac{2}{5}
\end{eqnarray}
where the last term in each line relates to the extra contribution
induced by the presence of the mirror doublet and singlet
bottom quarks, $n_g$ is the number of generations and $n_H$ is
the number of Higgs doublets. To be consistent with the electroweak
fits described earlier, 
we assume that only one Higgs doublet plays a relevant role 
in electroweak symmetry breaking.

This model predicts a shift of the hypercharge beta-function that
is somewhat smaller than the equal shifts of the beta-functions of the
weak and strong gauge couplings. 
As is well-known, within the SM, the strong and weak gauge couplings
meet at approximately $10^{17}$ GeV (for $n_H = 1$). The hypercharge 
coupling, however, crosses the other two at a much lower scale. 
This big `discrepancy' can be reduced by postulating a large number 
of Higgs doublets, but only at the cost  of bringing down the 
unification scale to $\sim 10^{12}$ GeV, a value palpably inconsistent
with proton stability. 
The introduction of the new quarks and the consequent 
shift in the beta-functions works to reduce this very same 
difference in the scales at which the couplings meet.
The improvement is quite significant. 
For $n_H = 1,2$ and 3, and `unification' scales
of approximately $5 \times 10^{16}$ GeV, $2 \times 10^{16}$ GeV
and $10^{16}$ GeV, the couplings differ from the average
\lq\lq unification\rq\rq value by less than 3, 1 and 3 percent
respectively\footnote{Two-loop effects will produce small modifications
	to these numbers.}.
These corrections are small, and considering 
the large scales involved, as emphasized above, could easily 
be accommodated (even for $n_H = 1$) by threshold
effects due to the presence of heavy particles (with GUT scale
masses) or Planck scale suppressed operators. 

Observe that, since the model lacks supersymmetry, dangerous
dimension five operators are absent from a potential
grand unified scenario. Moreover, the unification scale is
sufficiently large to avoid the constraints coming from proton decay
induced via dimension 6 operators. However, for the Higgs masses
and Yukawa couplings associated with the best fit to the
precision electroweak data, the model tends to induce a Landau
pole in the Higgs quartic couplings below the GUT scale,
particularly for relatively heavy quarks, $M_1 \simeq 250$ GeV. 
The most obvious means of avoiding this problem is to give 
up the property of perturbative unification. An alternate 
way would be to effect a suitable modification such as the one 
that we discuss shortly.

In the model with the non-standard mirror doublet quarks,
instead, the unification relations are not improved with
respect to the Standard Model case, since the shift in the
beta function of the hypercharge gauge coupling is larger than
in the ones associated with the strong and weak coupling
beta-functions. However, this model presents an interesting
property: the mirror doublet and singlet quarks introduced in this model
are contained in the adjoint ({\bf 24}), 
and in the {\bf 5} + ${\bf \bar{5}}$  
representations
of $SU(5)$.
Indeed, the weak doublet and singlet quarks in this model have precisely 
the same quantum numbers as the 24-plet partners of the standard
model gauginos and of the color-triplet 
Higgsinos of the minimal supersymmetric standard model (MSSM), 
respectively\footnote{In the minimal supersymmetric $SU(5)$ model, 
	the colored-triplet Higgsinos are assumed 
	to acquire GUT scale masses , creating the so-called
doublet triplet splitting problem. A similar hierarchy problem would
exist in the model described in this section.}.

  On a more speculative note, if one were willing to accept the presence
of a complete {\bf 24} of fermions at the weak scale, together
with the standard mirror doublet and singlet quarks\footnote{All these 
fields are contained in the 
	adjoint of $E_6$.}, one could obtain
excellent unification relations without the need of supersymmetry,
together with an excellent fit to the precision electroweak
observables. Within this assumption, the top-less mirror quark doublets
will be the ones leading to a relevant mixing with the bottom
quark, while the standard mirror quark doublets should have only
small mixing with the three generation of quarks. In this case, the
Higgs tends to be light. 
For a top-less doublet lighter than approximately 1 TeV, 
the Yukawa couplings are relatively weak and the Landau pole
problem is avoided.
This model may lead, instead, to a conflict with the Higgs potential
stability~\cite{Zhang:2000zy}. 
Whether this is 
a real physical problem can only be answered by studying the possibility
of ours being a metastable vacuum. In the SM with a similarly light
Higgs, $m_H \simeq 115$ GeV,
the requirement of strict stability would suggest the presence of new physics
far below the Planck scale. The requirement of being in a metastable
vacuum with lifetime longer than the age of the Universe, instead,
allows the Standard Model description to be valid up to scales close
to the Planck scale~\cite{Espinosa:1995se, Isidori:2001bm}. We postpone
for a future study a more detailed analysis of these questions.

Observe that the above-mentioned possibility leads to the potential
presence of fields with the quantum numbers of the MSSM gauginos
at low energies. 
In the absence of a symmetry
like R-Parity,  the fields with the quantum number of the Wino and
of the Bino will mix with the leptons and neutrinos, respectively,
and hence their couplings to the leptons and the Higgs bosons should
be very small. If one assumes the presence of a second Higgs doublet, with
no relevant role in the electroweak symmetry breaking mechanism, a 
coupling of order one of this field with the Bino-like field and, for
instance, the  third generation leptons may induce the proper annihilation 
rate to make the Bino-like field a good dark matter 
candidate\footnote{Discrete symmetries may need to be imposed
	in order to avoid dangerous lepton flavor violating processes.}. 
Alternatively,  
the Bino may play the role of a sterile neutrino.
Without the addition of new fields, the gluino-like particle tends
to be very long lived or even stable. 
We also reserve for a separate study the analysis of the cosmological
and phenomenological consequences of such a scenario.

\section{Conclusions}

The Standard Model with a light Higgs boson is in very good
agreement with the precision electroweak observables measured
at the Tevatron, SLD and LEP colliders. Although there is no
clear indication of the need for new physics in the electroweak
precision measurement data, the prediction for the effective
leptonic weak mixing angle extracted from the hadronic and
leptonic observables are several standard deviations away
from each other. In this article we have analyzed a possible
way of fixing this discrepancy by introducing mirror quarks
with quantum numbers similar to those of the left-handed
and right-handed bottom quarks.

While the two models analyzed in our article lead to an improvement of 
the general fit to the precision electroweak data, they
present qualitatively different characteristic that make themselves 
easily distinguishable
from the experimental point of view. In the model with standard
mirror quark doublets, only negative shifts to the right-handed
bottom coupling may be obtained by means of the mixing with
the doublet and singlet quarks. The very smallness of this
coupling within the SM, however, 
allows us not only to change its magnitude but reverse 
its sign as well. Apart from improving the agreement to the 
precision electroweak data to a great extent, this also leads 
to interesting predictions for the bottom
quark asymmetries away from the $Z$ peak. Moreover, the best fit
to the data within this model is obtained for quarks light enough 
to be accessible at the Tevatron collider, and relatively heavy Higgs bosons.
Finally, the unification relations are significantly improved
with respect to the Standard Model case, and the potential unification
scale is  sufficiently large in order to avoid proton decay via
dimension six operators. However, perturbative unification within this
simple extension of the Standard Model is not
possible, due to the presence of a Landau pole in the Higgs quartic
couplings at scales below the potential GUT scale.

On the other hand, the model with non-standard mirror quark doublets
leads to mild modifications of the left- and right-handed bottom
quark couplings induced via small mixings of the mirror quarks with
the standard ones. Besides this, the model
prefers relatively light Higgs bosons, possibly in the range
testable at the next run of the Tevatron collider, while the
mirror quarks tend to be heavy, only accessible at the LHC.
An interesting property of this model is that the quantum numbers
of the exotic quarks are precisely the ones of the heavy coloured gauginos
and Higgsinos within minimal supersymmetric SU(5) scenarios. 

One can contemplate the possibility of taking both standard and exotic 
mirror quark doublets and the mirror down quark singlets, and  
including particles with the quantum numbers of the standard
gauginos in the minimal supersymmetric Standard Model and,
eventually, an additional Higgs doublet. This extension of the 
Standard Model allows a remarkable improvement in the fit to the
precision electroweak observable data, leads to
the possibility of achieving consistency with the unification
of gauge couplings and has all the ingredients necessary to lead to
an explanation of the dark matter content of the Universe. 

\vspace{1.cm}
~\\
{\Large \bf Acknowledgements}\\
~\\
We would like to thank T. LeCompte, P.Q. Hung and C.-W. Chiang 
for useful discussions.
CW would also like to thank M.~Carena, J.~Erler and M.~Peskin for 
useful comments. Work supported in part by the US DOE, Div.\ of HEP,
Contract W-31-109-ENG-38. DC thanks the Deptt. of Science and 
Technology, India for financial assistance under the 
Swarnajayanti Fellowship grant.

\end{document}